\documentclass[10pt,letterpaper]{article}

\usepackage{ccn}
\usepackage{pslatex}
\usepackage{apacite}
\usepackage{amsmath}  
\usepackage{amsfonts}
\usepackage{rotating}
\usepackage{adjustbox}
\usepackage[font=small,labelfont=bf]{caption}

\title{Constructing and deconstructing bias: modeling privilege and mentorship in agent-based simulations}

\author{{\large \bf Andria L. Smith$^*$ (asmith@is.mpg.de)} \\
  Max Planck Institute for Intelligent Systems, Heisenbergstr. 3\\
Stuttgart, Baden-Württemberg 70569 Germany
  \AND {\large \bf Simon Heuschkel$^*$ (simon.heuschkel@student.uni-tuebingen.de)} \\
  University of Tübingen, Maria-von-Linden-Str. 6\\
Tübingen, Baden-Württemberg 72076 Germany
\AND {\large \bf Ksenia Keplinger (kkeplinger@is.mpg.de)} \\
  Max Planck Institute for Intelligent Systems, Heisenbergstr. 3\\
Stuttgart, Baden-Württemberg 70569 Germany
\AND {\large \bf Charley M. Wu (charley.wu@uni-tuebingen.de)} \\
  University of Tübingen, Maria-von-Linden-Str. 6\\
Tübingen, Baden-Württemberg 72076 Germany
\AND $^*$ These authors contributed equally
}

\begin{document}

\maketitle 


\section{Abstract}

{
Bias exists in how we pick leaders, who we perceive as being influential, and who we interact with, not only in society, but in organizational contexts. Drawing from leadership emergence and social influence theories, we investigate potential interventions that support diverse leaders. Using agent-based simulations, we model a collective search process on a fitness landscape. Agents combine individual and social learning, and are represented as a feature vector blending relevant (e.g., individual learning characteristics) and irrelevant (e.g., race or gender) features. Agents use rational principles of learning to estimate feature weights on the basis of performance predictions, which are used to dynamically define social influence in their network. We show how biases arise based on historic privilege, but can be drastically reduced through the use of an intervention (e.g. mentorship). This work provides important insights into the cognitive mechanisms underlying bias construction and deconstruction, while pointing towards real-world interventions to be tested in future empirical work. 

}
\begin{quote}
\small
\textbf{Keywords:} 
social influence; bias; privilege; social network; intervention
\end{quote}

\section{Introduction}
Bias exists in how we pick leaders, who we are influenced by, and who we interact with. For instance, there are more CEOs named John or David than women CEOs in the S\&P 1500 companies \cite{johnson2016if}. Despite increased interest in creating more diverse and inclusive organizational environments, there are many barriers in place, such as biases, preventing progress \cite{keplinger2022stigmatization}.

Although empirical research on the sources of biases and potential interventions for unlearning biases has a long tradition \cite{axelrod1997dissemination, freeman2011looking,serban2015leadership,schelling1971dynamic}, it is still unclear when, why, and how privilege and bias arise \cite{colella2017one}. Specifically, we are interested in how privilege, defined as unearned access to rewards and resources for specific groups \cite{case2012systems,crevani2019privilege}, hinders the emergence of marginalized leaders \cite{badura2022leadership}.  
Here, to integrate theories on leader emergence and social influence, we use agent-based simulations which are a computational approach still rarely applied to the leadership context \cite<but see>[]{cao2020agent}. These simulations add precision to previous verbal theories \cite{samuelsoncomputational,vancouver2020translating} and shed light on the cognitive mechanisms underlying the learning of biases towards arbitrary agent features (e.g., race, gender, age, etc.).

This study aims to 1) demonstrate how biases are recreated through rational principles of multi-agent learning when certain agents are placed in privileged locations in the environment and 2) investigate the impact of an intervention, where we create temporary social network connections between high and low performing agents modeled as external agents (e.g., mentors or role models), to unlearn the bias. Our simulation shows how we can systematically reduce bias across all agents, thus leading to an increase in diverse representation of emergent leaders.


\begin{figure}[t!]
\centering
\includegraphics[width=\columnwidth]{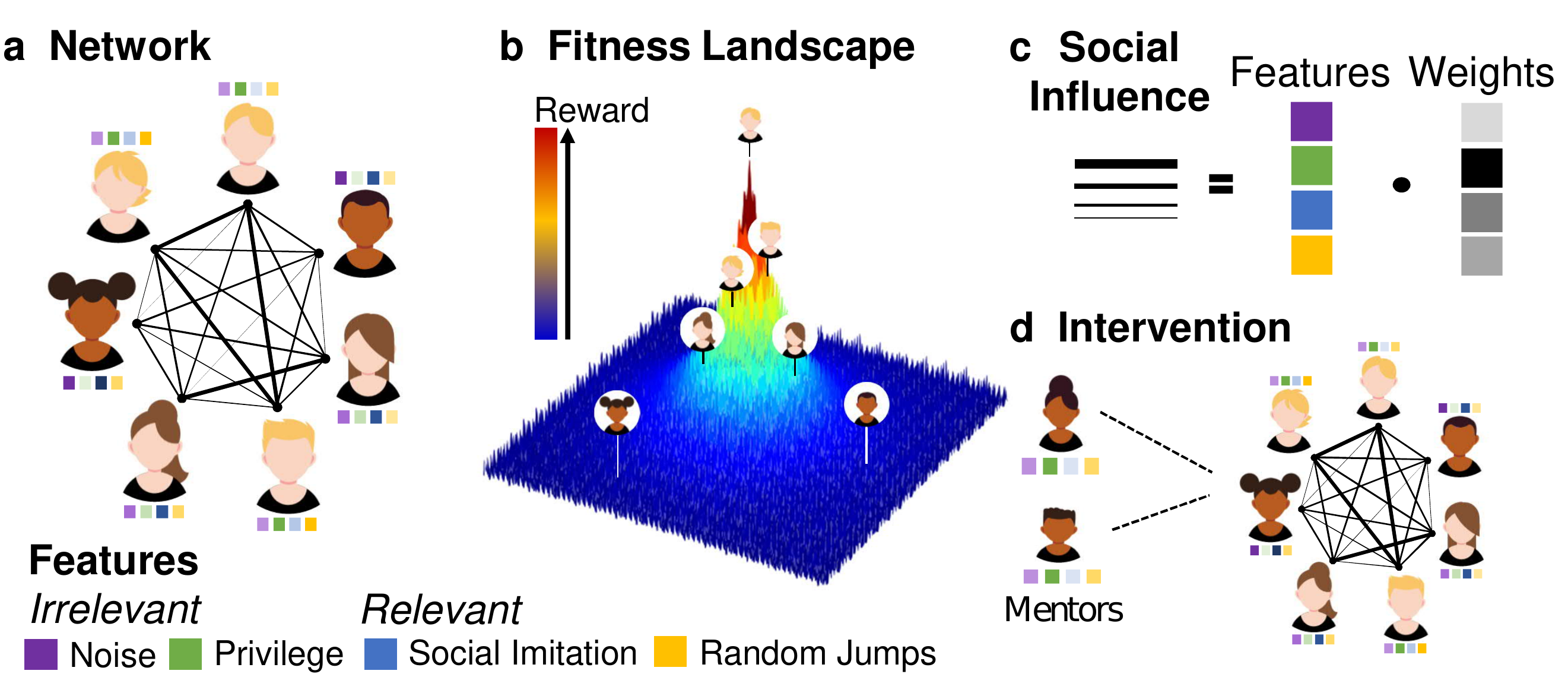}
\caption[]{Agent-based simulations. \textbf{a}) Network structure and agent features. \textbf{b}) Fitness landscape. \textbf{c}) Social influence learning as function of learned feature weights. \textbf{d}) Intervention through mentorship.}
\label{fig:task}
\end{figure} 

\section{Methods}

We use agent-based simulations, where a team of 7 agents collectively optimize a two-dimensional fitness landscape with multiple local optima. We use the Ackley and Drop Wave environments as common test functions for optimization algorithms \cite{simulationlib}, in addition to the \citeA{MasonWatts} environment, with previous work characterizing these environments as having similar average payoff and number of local optima \cite{barkoczi2016collective}. We define each landscape with $1000 \cdot 1000$ discrete locations. 


Agents are connected by a weighted social influence matrix $A$ and defined by a set of features (Fig.~\ref{fig:task}a). On each iteration, agents update their position on the fitness landscape $\mathbf{x}_i$ and the according fitness value $y_i$ using a search policy combining individual and social learning (Fig.~\ref{fig:task}b), and then update social influence based on learned feature weights $W$ (Fig.~\ref{fig:task}c).

\subsubsection{Social influence.}
Each agent $i$ is defined by a set of features $\mathbf{f}_i = [\gamma, \eta, \rho, \nu]$ representing personal attributes that are \textit{policy relevant} (i.e., capturing characteristics of learning strategy: $\gamma$, $\eta$), indicate \textit{privilege} (i.e., starting condition $\rho$) or are completely \textit{irrelevant} (i.e., noise $\nu$). 

At every iteration each agent tries to estimate the performance of other agents $j$ with a linear weighted sum of their features $\hat{r}_{i,j} = \mathbf{w}_i^\top \mathbf{f}_j + \epsilon$, where $\epsilon=\min_j r_j$ is an offset for the lowest current reward across all agents in the group. The performance of other agents is measured using temporally discounted past rewards relative to the current time point $T$: $r_{i, T} = (\sum^T_{t=0} y_{i,t} \cdot \lambda^{T-t})/ \sum^T_{t=0} \lambda^{t}$, where we set the temporal discount $\lambda = 0.9$. 
Weights are updated by minimizing the mean squared prediction error between the actual rewards and their predictions through gradient descent:
\begin{equation}
\label{eq:SGD}
    \mathbf{w}_i \leftarrow \mathbf{w}_i - \alpha \cdot  \frac{\partial}{\partial \mathbf{w}_i} \mathcal{L}_{MSE}(\mathbf{r}, \mathbf{\hat{r}}_{i})
\end{equation}
with learning rate $\alpha = 0.1$.

Thus, the learning of weights captures feature-specific biases of social influence, where agents with highly weighted features will exert more social influence. We update the social influence matrix every iteration as a function of predicted performance: $A \leftarrow A + \beta \cdot W \cdot F^\top$ with $\beta=0.5$ controlling the update rate. 

\subsubsection{Individual and social learning.}
Agents use a combination of social and individual learning policies to optimize their position in the fitness landscape. 
Each agent $i$ first uses social imitation with probability  $P(\gamma_i)$, whereby it uses a softmax imitation policy as a function of social influence weights:
\begin{equation}
    \pi_{\textrm{imitation}}(\mathbf{x}_j) \propto exp(\frac{a_{i,j}}{\sum_k a_{i,k}} / \tau)
    \label{eq:imitation}
\end{equation}
\noindent Intuitively, agents are more likely to imitate others with higher perceived influence $a_{i,j}$. 

If the social policy is not enacted, with probability $1-P(\gamma_i)$, the agent uses individual learning. First, the agent tries a random jump with a probability of $P(\eta_i)$, whereby it evaluates a random position in the landscape and jumps there if it increases its fitness value. If the agent does not use a random jump, it performs local optimization using stochastic hill climbing (SHC) over all neighbouring positions: 
\begin{equation}
    \pi_{\textrm{SHC}}(\mathbf{x}') \propto \exp(y_i'/\tau),
    \label{eq:SHC}
\end{equation}
\noindent where each $y_i$ is the fitness value of a neighboring solution $\mathbf{x}'$, and with higher-valued solutions more likely to be selected. In all cases, we set $\tau = 0.01$. 

Policy relevant features $\gamma$ and $\eta$ are sampled uniformly from $\mathcal{U}(0,0.1)$, whereas privilege $\rho$ and noise $\nu$ are sampled from a multimodal Gaussian with two different means $\in (0.03,0.07)$, each with variance of 0.01, and truncated  between $[0,0.1]$. 
Both policy relevant features influence performance by aiding in escaping local optima, but $\gamma$ depends on the quality of social information.  

\subsubsection{Privilege and intervention.}  
To model \textit{privilege}, we use $\rho$ to define the starting position of an agent, by placing them in a location within the top $(\rho \times 10)$-th quantile $\pm 0.005$ of rewards. This initialization leads to a positive correlation between privilege and performance at the beginning of the simulation (e.g., $\rho=0.5$ will start near the median reward in the landscape), but has no bearing on  learning capabilities.

We hypothesize that agents will develop a bias towards learning large feature weights for privilege. Therefore, we introduce an intervention in form of \textit{mentorship}. Mentors are modeled as agents who have less privilege $\rho \sim \mathcal{N}(0.03,\,0.01)$, but have high policy relevant traits $\eta, \gamma \sim \mathcal{N}(0.08,\,.005) \in [0,0.1]$ and high performance $r_{\textrm{mentor}} > 90\% \; \textrm{of all  fitness values}$. At each iteration, a less privileged agent ($\rho < .05$) gets a mentor assigned with probability 0.2. The features and $r_{\textrm{mentor}}$ of this mentor are used to optimize the social weight of the mentee $\mathbf{w}_i$, additionally to the group members.
Thus, mentors do not participate in the collective optimization and therefore are not targets for social imitation or part of the social influence network, but only support social feature learning (Eq.~\ref{eq:SGD}).
However, mentors signal awareness of noise features to the group that certain biases can be broken \cite{freeman2019exploring,ivey2022workplace,raza2020reverse,williams2020peer}.

\section{Results}
\begin{figure}[t!]
\centering
\includegraphics[width=\columnwidth]{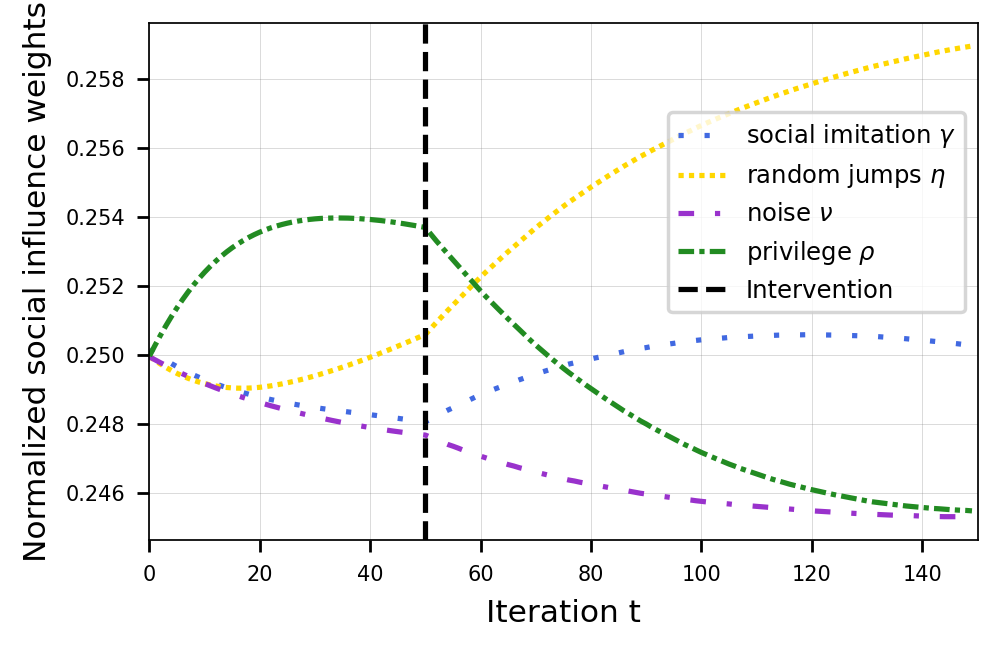}
\vspace{-.45cm}
\caption[Results.]{Results. Impact of the intervention on the mean feature weights across fitness landscapes. The normalized weights indicate how much performance is credited to each feature.}
\label{fig:result}
\end{figure} 


We ran 1000 simulations with 150 iterations per environment. Figure \ref{fig:result} shows that even though agents start out unbiased, by iteration 25 they learn strong weights for the privilege feature $\rho$. 
As the simulation continues, random jumps $\eta$ prove to be useful and the agents learn increasingly strong weights for this feature. 
Social imitation in form of $\gamma$ does not seem to be valued initially, with decreasing weights until after the intervention, which may be due to a masking effect of the strong privilege weights. 

After the intervention (vertical dashed line), privilege weights decrease strongly, while social imitation weights begin to increase. This shows that mentors do not only reduce the influence of policy irrelevant features (i.e., privilege), but also help agents learn to rely more on policy relevant features $\gamma$ and $\eta$.  
Although privilege and noise weights decrease faster after the intervention, agents still rely more on privilege $\rho$ than noise $\nu$, showing how difficult it is to get fully rid of a learned bias.
Simulations without the intervention also result in decay in privilege weights that is much less pronounced and fails to encourage increasing weights for social imitation.

\section{Conclusion}  



Our computational approach provides a tool for understanding how rational principles of learning can shape the formation of biases in which features are assigned credit for performance. We show how biases naturally arise based on historic privilege, but can be mitigated through an intervention by creating mentoring relationships between high and low performing marginalized agents. This work provides important insights into the cognitive mechanisms underlying how biases can develop and be unlearned, while pointing towards real-world interventions to be tested in future empirical work. 



\section{Acknowledgements}
This work is supported by the German Federal Ministry of Education and Research (BMBF): Tübingen AI Center, FKZ: 01IS18039A and funded by the Deutsche Forschungsgemeinschaft (DFG, German Research Foundation) under Germany’s Excellence Strategy–EXC2064/1–390727645.

\bibliographystyle{apacite}

\setlength{\bibleftmargin}{.125in}
\setlength{\bibindent}{-\bibleftmargin}

\bibliography{ccn}

\end{document}